# Empirical Study of Deep Learning for Text Classification in Legal Document Review

Fusheng Wei
Advanced Analytics
Ankura
Washington, D.C. USA
fusheng.wei@ankura.com

Han Qin
Advanced Analytics
Ankura
Washington, D.C. USA
han.qin@ankura.com

Shi Ye
Advanced Analytics
Ankura
Washington, D.C. USA
shi.ye@ankura.com

Haozhen Zhao
Advanced Analytics
Ankura
Washington, D.C. USA
haozhen.zhao@ankura.com

*Abstract*— **Predictive coding has been widely used in legal matters to find relevant or privileged documents in large sets of electronically stored information. It saves the time and cost significantly. Logistic Regression (LR) and Support Vector Machines (SVM) are two popular machine learning algorithms used in predictive coding. Recently, deep learning received a lot of attentions in many industries. This paper reports our preliminary studies in using deep learning in legal document review. Specifically, we conducted experiments to compare deep learning results with results obtained using a SVM algorithm on the four datasets of real legal matters. Our results showed that CNN performed better with larger volume of training dataset and should be a fit method in the text classification in legal industry.**

## I. INTRODUCTION

In the legal industry, due to rapidly growing volume of electronically stored information, the costs involved in manually reviewing an overwhelming number of documents have grown dramatically. Companies regularly spend millions of dollars producing responsive documents [1]. To more efficiently cull through massive volumes of data for relevant information, attorneys have been using text classification, a supervised machine learning technique typically referred to as predictive coding or technology assisted review (TAR) in the legal domain.

Traditionally, Logistic Regression (LR) and Support Vector Machines (SVM) have been two popular machine learning algorithms used in predictive coding. Studies have been carried out in legal domain to better understand the underlying techniques such as preprocessing parameters [2] and active learning [3] to make predictive coding more effective.

In recent years, deep learning has made tremendous progress for machine learning and AI. Since the breakthroughs of using neural network in visual analysis and natural language processing including speech recognition and language translation, deep learning technique such as convolution network has been adapted to text classification and has been approved effective in academic researches and demonstrated with real world data such as Yelp and IMDB reviews for predicting customer ratings and Tweet messages for sentiment analysis. Convolution network is also powerful in feature extraction and it has the capability of preserving word order due to its sequence-based nature.

As predictive coding being widely used in legal industry and deep learning is showing its promises in wide range of vertical domains, it is natural to ask how deep learning performs for predictive coding, that is, how deep learning works for text classification in legal domain.

In this paper, we carry out experiments to do an empirical study of deep learning, particularly convolution network, for text classification using real data collected from various legal projects. We apply the architecture to text classification problem in the legal domain with datasets of documents collected from various projects on real legal cases. Our goal of this study is first to experiment the effectiveness of deep learning method in classification and then compare the effectiveness with that of the SVM method.

## II. DEEP LEARNING FOR TEXT CLASSIFICATION

### A. Newral Network Architecuures

A variety of neural network architecture have been exploited for text classification. The simplest approach could be taking the feature input from a linear model such as LR or SVM and feed them into a Deep Neural Network (DNN), which serves a nonlinear learner to replace the linear ones in traditional machine learning methods.

Convolution Neural Network (CNN) has been shown as a powerful tool in image analytics, especially for its use for feature extraction with transfer learning. CNN has been adapted to text classification and showed to be useful for classification tasks in which we expect to find strong local clues regarding class membership, such as a few key sentences or phrases. In a CNN for text classification, convolution layers involve one dimensional convolution with a small size kernel to extract features, and max pooling to condense or summarize the features extracted from the convolution. Finally, the fully connected layer takes the features through activations and fit to the training data and make predictions. This is the architecture used for this paper for the empirical study. Detail settings will be described in the next section.

Other architectures include the use of recurrent neural network and recursive neural network, both are denoted as RNN. For more comprehensive history on text classification using RNN and other architectures above, one can refer to [4].

### B. Word Embedding

Regardless of which neural network architecture to choose, it all starts with word representation as inputs to the neural network. In neural network for text classification, word embedding is often used to perform the task. Word embedding is to represent each word as a vector in a low dimensional space. Contrast to one-hot vectors for the representation, which involves large sparse representation matrix, word embedding maps words to vectors of fixed length, say, 100. While with one-hot vectors each word is represented independent of other words, word embedding is a representation of words where different words having similar meaning also have a similar representation.

Word embedding can be trained as supervised learning in the first layer of the neural network, or it can be unsupervised trained beforehand. Two popular embedding by unsupervised learning are word2vec and GloVe, we used the latter for this study as well as supervised trained as part of the embedding layer of the neural network model.

## III. EXPERIMENTAL SETTINGS AND DATA SETS

### A. Model setup

In this paper we use a common architecture that consists of a word embedding, a convolutional model, and a fully connected model. We use Keras (Tensorflow backend) to implement the model, the summary of the model is shown below in Table 1:

*Table 1. Keras model summary*

| Layer (Type) | Output Shape | Param # |
| --- | --- | --- |
| embedding_1 (Embedding) | (None, 1500, 100) | 2000000 |
| dropout_1 (Dropout) | (None, 1500, 100) | 0 |
| conv1d_1 (Conv1D) | (None, 1496, 64) | 32064 |
| max_pooling1d_1 (MaxPooling1 | (None, 374, 64) | 0 |
| flatten_1 (Flatten) | (None, 23936) | 0 |
| dense_1 (Dense) | (None, 1) | 23937 |

The word embedding can use either self-trained together with the whole sequence model or use pre-trained representations such as Glove or Word2Vec as transfer learning.

### B. Dataset

We use four different datasets in real legal matters, named projects A, B, C, D, respectively, each of them contains millions of records, with a large number of the samples labeled. For each data set, we first set aside a randomly selected set of labeled records with size around 25000 as the test set. Then we generate four incremental training sets by randomly selecting from the remaining labeled records with certain proportions. Thus, the amounts of training sets are different due to the original volumes of data in different projects. Table below shows the configurations:

*Table 2. Details of Training and Test datasets for different projects*

| Projects | Label | Test Set | Train_Set1 | Train_Set2 | Train_Set3 | Train_Set4 |
| --- | --- | --- | --- | --- | --- | --- |
| A | NEG | 20,927 | 2,811 | 5,526 | 13,958 | 28,072 |
| A | POS | 4,073 | 533 | 1,163 | 2,765 | 5,374 |
| A | Ratio | 16.29% | 15.94% | 17.39% | 16.53% | 16.07% |
| B | NEG | 18,801 | 4,850 | 9,679 | 19,403 | 48,392 |
| B | POS | 5,414 | 1,378 | 2,818 | 5,447 | 13,763 |
| B | Ratio | 22.36% | 22.13% | 22.55% | 21.92% | 22.14% |
| C | NEG | 21,698 | 1,029 | 2,062 | 4,135 | 10,399 |
| C | POS | 3,302 | 164 | 325 | 639 | 1,536 |
| C | Ratio | 13.21% | 13.75% | 13.62% | 13.39% | 12.87% |
| D | NEG | 14,593 | 1,529 | 3,031 | 6,069 | 15,028 |
| D | POS | 9,730 | 977 | 1,979 | 3,952 | 10,037 |
| D | Ratio | 40.00% | 38.99% | 39.50% | 39.44% | 40.04% |

We train the deep learning model on each of the training sets and then test it with the testing set for each increment, and then do the same with SVM. Precision and recall curves are plotted as the performance metrics for each setting.

For each of the experiments, we first load the text data and then apply basic cleaning for the loaded texts, which include dropping stop words, changing to lower case, and dropping numbers off words. This is to shorten the lengths of texts before tokenizing them for input to the neural network. For all the sample sizes, we found an upper limit of 1500 words are a good number for the cleaned text, with the following histogram typical across all samples (Figure 1). We therefore choose this length as the cutoff when mapping the tokenized text to sequences, with padding zeros for shorter text.

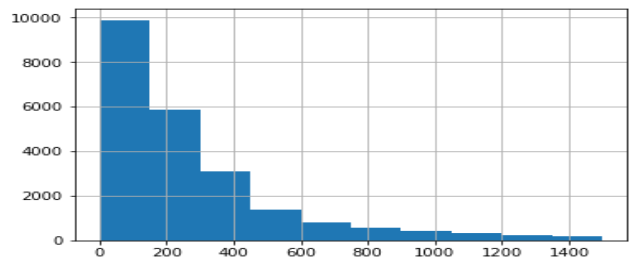

*Figure 1. Histogram of the sizes of a typical sample set*

## IV. RESUTLS

The classification accuracies obtained from SVM and CNN models which are trained on four training sets of four projects are shown in Table 3.

*Table 4. Classification accuracies of SVM and CNN models*

| Project | Method | Train_Set1 | Train_Set2 | Train_Set3 | Train_Set4 |
|---|---|---|---|---|---|
| A | SVM | 86.14% | 85.92% | 85.45% | 86.66% |
| A | CNN | 84.57% | 84.85% | 85.21% | 85.35% |
| B | SVM | 77.09% | 76.96% | 75.67% | 78.40% |
| B | CNN | 78.35% | 79.35% | 80.09% | 80.21% |
| C | SVM | 76.57% | 78.83% | 78.43% | 81.59% |
| C | CNN | 78.60% | 79.34% | 81.31% | 83.66% |
| D | SVM | 91.67% | 92.53% | 92.49% | 91.89% |
| D | CNN | 91.04% | 91.58% | 91.76% | 92.89% |

### A. CNN Results

In this paper, we used 0.5 as the threshold value to classify positive and negative samples and used a large test dataset (around 25,000 samples) for each project. The classification accuracy on the same test data with different training data changes dynamically in each project. The results demonstrated that the larger volume of training data produces better performances with this CNN model, whereas the smaller volume of training data may perform worse (Table 3). In the experiments of all four projects, the classification accuracies are high, and the accuracy rate consistently rises as the volume of training dataset increases. Training dataset in large volumes are all over 0.8, which indicates that CNN should be a fit method in the text classification in legal industry. However, compare among different projects, the test accuracy in Project B reaches 0.8 and the test accuracy in Project D towards ~0.93. The reason could be the total amount of training set in these two projects - there are 62,155 samples in Project B and 11,935 samples in Project D. This does not indicate the larger volume of training sample in different projects performs worse test accuracy, because Project A and C have higher accuracy score than Project B while both have more training samples. Thus, we assume that the data quality and the ratio of positive to negative samples also effect the test accuracies, which will be discussed in further experiments.

### B. Comparative studies between CNN classifier and SVM classifier

The trends of precisions of the CNN classifier for the four projects A/B/C/D are similar to the trends that performed with the SVM classification method. However, the precisions with CNN classifier are higher than the precisions with SVM classifier while using larger volume of training dataset. Figure 2 to 4 represent the differences between CNN and SVM models in different projects. It is evident that the CNN model outperforms SVM with a large volume of training dataset. With respect to Precision in the 75% recall in all projects, excluding the values in Project A and B because of the abnormal low precision values, we find that the CNN model has higher values than SVM models in Project C (82% to 78%) and D (68% to 60%). These findings state clearly that CNN outperforms traditional text mining approaches for text classification presenting the potential for further development on binary text identification in legal industry.

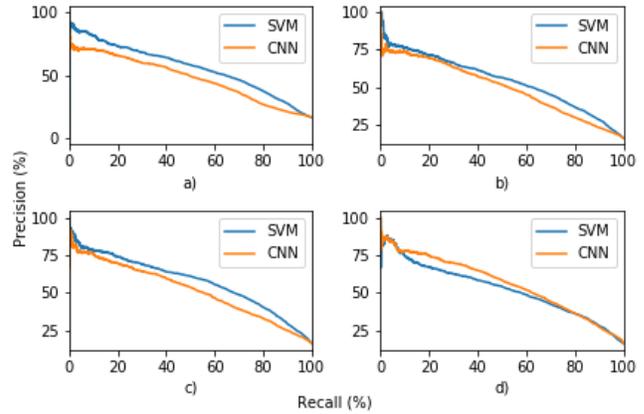

*Figure 2. Precision and recall curve of project A.*
*a) – d) results of models trained on different training datasets (smallest to largest)*

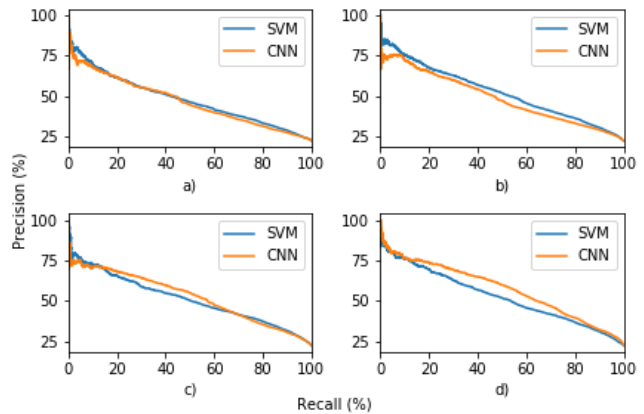

*Figure 3. Precision and recall curve of project B.*
*a) – d) results of models trained on different training datasets (smallest to largest)*

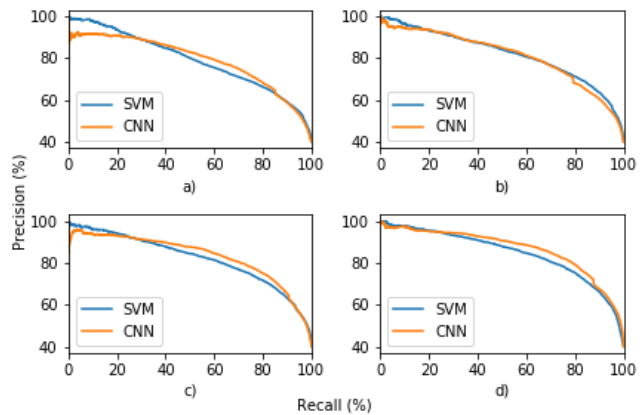

*Figure 4. Precision and recall curve of project C.*

*a) – d) results of models trained on different training datasets (smallest to largest)*

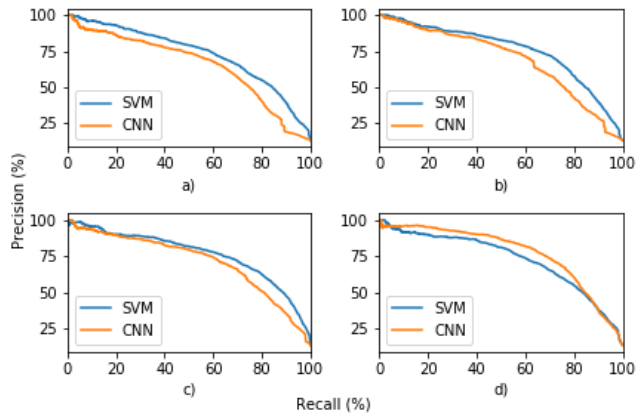

*Figure 5. Precision and recall curve of project D.*
*a) – d) results of models trained on different training datasets (smallest to largest)*

## V. CONCLUSION AND FUTURE WORK

This research examines the capability of a deep learning model based on CNN for binary classification (responsive and non-responsive) in real legal matters. In this paper, a basic CNN model is used in the experiments without further optimized, the results still show a higher performance in direct comparison to traditional approach – SVM on the larger dataset and a more stable growth trend with the gradually increasing amount of training samples.

While convolution neural network provides an effective way of text classification by learning the text in sequences compared to bag of word methods that lacks the sequence information, thus better extract features in terms of sentences/phrases, the challenge lies in the actual identification of relevant sentence(s), as legal documents are often identified as relevant due to a few sentences or short passages. This is directly related to explainable predict analysis. A paper in this conference by the Ankura group presented a method. We would like to research on how that approach can be related to deep learning method described here. More specifically, the work in [6] has a large set of annotated sentences as part of the labeled text. So, training those sentences together with full text documents may provide some information on the relevant sentences.

The other challenge is the sequence approach has a limit of the sequence length, so for large text, how can one keep the relevant part of the text in the training without chopping off before feeding to the training. In the Google Development Guide, it introduced a threshold of S/W, where S stands for the number of samples and W the median number of words in a text, and the threshold is used to make choice of machine learning methods, that is, if S/W < 1500, use traditional methods, otherwise use deep learning. It would be interesting to examine our experiments against Google's guideline.

For word embedding, experiments in this paper shows that training the word representation weights as part of overall training (self-trained word imbedding) outperforms pretrained Glove word embedding. One reason of the difference can be the size of the training samples, the other can be that Glove is too general with respect to legal domain. Therefore, it is interesting to exploit a word embedding for the latter.

The experiments showed that the CNN models take long time to train. With our analytic tools based on traditional methods, users can select a training set and generate a model within a few minutes. To enhance the deep learning approach to such a level we would need GPU to speed up the training and predicting processes, so when we incorporate deep learning to the current tool as an alternative modeling option for the users, we'll need to consider adding GPU to the configuration.